\documentclass[twocolumn]{bmcart}

\usepackage{amsthm,amsmath}
\usepackage{amssymb,bm}
\RequirePackage{natbib}
\RequirePackage{hyperref}
\usepackage{soul,color}
\usepackage{dblfloatfix}

\soulregister\cite7
\soulregister\ref7
\soulregister\pageref7

\usepackage{graphicx}

\startlocaldefs
\endlocaldefs

\begin{document}

\begin{frontmatter}

\begin{fmbox}
\dochead{Review}


\title{Ubiquitous cell-free Massive MIMO communications}


\author[
   addressref={aff1},                   
   corref={aff1},                       
   email={giovanni.interdonato@liu.se}   
]{\inits{GI}\fnm{Giovanni} \snm{Interdonato}}
\author[
   addressref={aff1},
   email={emil.bjornson@liu.se}
]{\inits{EB}\fnm{Emil} \snm{Bj\"{o}rnson}}
\author[
   addressref={aff3},
   email={hien.ngo@qub.ac.uk}
]{\inits{HQN}\fnm{Hien Quoc} \snm{Ngo}}
\author[
   addressref={aff2},
   email={pal.frenger@ericsson.com}
]{\inits{PF}\fnm{P\aa l} \snm{Frenger}}
\author[
   addressref={aff1},
   email={erik.g.larsson@liu.se}
]{\inits{EGL}\fnm{Erik G} \snm{Larsson}}


\address[id=aff1]{
  \orgname{Department of Electrical Engineering (ISY), Link\"{o}ping University}, 
  \postcode{581 83}                               
  \city{Link\"{o}ping},                           
  \cny{Sweden}                                    
}
\address[id=aff2]{%
  \orgname{Ericsson Research, Ericsson AB},
  \postcode{583 30}
  \city{Link\"{o}ping},
  \cny{Sweden}
}
\address[id=aff3]{%
  \orgname{Electrical Engineering and Computer Science (EEECS), Queen's University Belfast},
  \postcode{BT3 9DT}
  \city{Belfast},
  \cny{U.K.}
}
  

\begin{artnotes}
\end{artnotes}



\begin{abstractbox}

\begin{abstract} 
Since the first cellular networks were trialled in the 1970s, we have witnessed an incredible wireless revolution. From 1G to 4G, the massive traffic growth has been managed by a combination of wider bandwidths, refined radio interfaces, and network densification, namely increasing the number of antennas per site. Due its cost-efficiency, the latter has contributed the most. 
Massive MIMO (multiple-input multiple-output) is a key 5G technology that uses massive antenna arrays to provide a very high beamforming gain and spatially multiplexing of users, and hence, increases the spectral and energy efficiency.
It constitutes a centralized solution to densify a network, and its performance is limited by the inter-cell interference inherent in its cell-centric design. Conversely, ubiquitous cell-free Massive MIMO refers to a distributed Massive MIMO system implementing coherent user-centric transmission to overcome the inter-cell interference limitation in cellular networks and provide additional macro-diversity. These features, combined with the system scalability inherent in the Massive MIMO design, distinguishes ubiquitous cell-free Massive MIMO from prior coordinated distributed wireless systems. In this article, we investigate the enormous potential of this promising technology while addressing practical deployment issues to deal with the increased back/front-hauling overhead deriving from the signal co-processing.
\end{abstract}




\end{abstractbox}
\end{fmbox}

\end{frontmatter}




\section{\normalsize Introduction} \label{Sec:Introduction}

One of the primary ways to provide high per-user data rates–-requirement for the creation of a 5G network---is through network densification, namely increasing the number of antennas per site and deploying smaller and smaller cells~\cite{Andrews2017a}. A communication technology that involves base stations (BSs) with very large number of transmitting/receiving antennas is Massive MIMO~\cite{Marzetta2016a}, where MIMO stands for multiple-input multiple-output. This key 5G technology leverages aggressive spatial multiplexing. 
In the uplink (UL), all the users transmit data to the BS in the same time-frequency resources. The BS exploits the massive number of channel observations to apply linear receive combining, which discriminates
the desired signal from the interfering signals. 
In the downlink (DL), the users are coherently served by all the antennas, in the same time-frequency resources but separated in the spatial domain by receiving very directive signals. By supporting such a highly spatially-focused transmission (precoding), Massive MIMO provides higher spectral efficiency, and reduces the inter-cell interference compared to existing mobile systems.   

\begin{figure*}[!t]\centering
	\includegraphics[width=.98\textwidth]{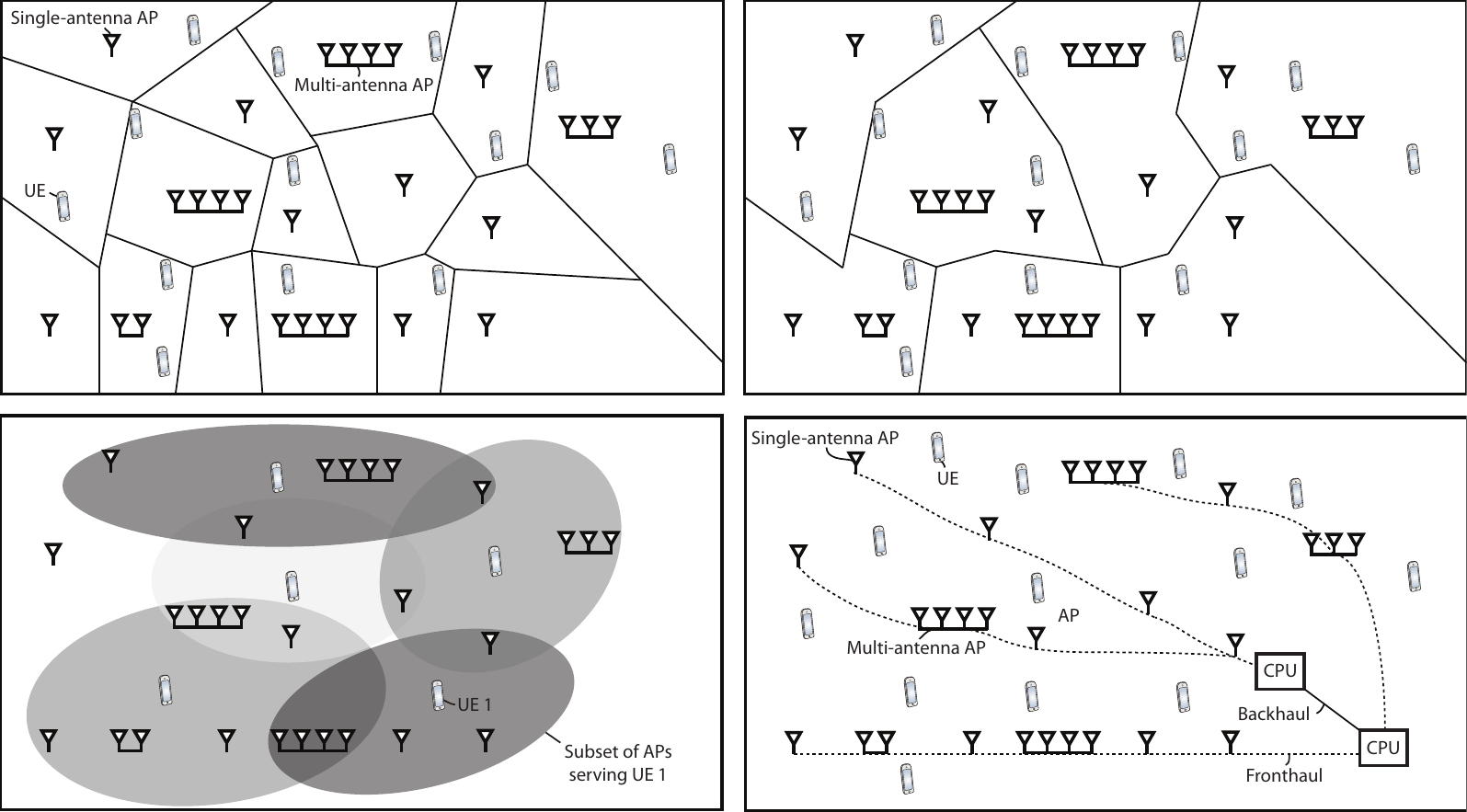}
  	\caption{\csentence{Example of network deployments.}
      Top-left: A conventional cellular network where each UE is connected to only one AP. Top-right: A conventional network-centric implementation of CoMP-JT, where the APs in a cluster cooperate to serve the UEs residing in their joint coverage area. Bottom-left: A user-centric implementation of CoMP-JT, where each UE communicates with its closest APs. Bottom-right: A ``cell-free'' Massive MIMO network is a way to implement a user-centric network.}
  	\label{fig:figure1}      
\end{figure*}

The inter-cell interference is however becoming the major bottleneck as we densify the networks. It cannot be removed as long as we rely on a network-centric (cell-centric) implementation, since the inter-cell interference is inherent to the cellular paradigm~\cite{Lozano2013a}. In a conventional cellular network, each user equipment (UE) is connected to the access point (AP) in only one of the many cells (except during handover). At a given time instance, the APs have different numbers of active UEs, causing inter-cell interference (Fig.~\ref{fig:figure1}, top-left). 

Cellular networks are suboptimal from a channel capacity viewpoint because higher spectral efficiency (SE) (bit/s/Hz/user) can be achieved by co-processing each signal at multiple APs \cite{Shamai2001a}. 
The signal co-processing concept is present in~\cite{Zhou2003a}, network MIMO~\cite{Venkatesan2007a,Caire2010b}, coordinated multipoint with joint transmission (CoMP-JT) \cite{Boldi2011a,Marsch2011a,irmer2011coordinated}, and multi-cell MIMO cooperative networks \cite{Gesbert2010a}. 
It is conventionally implemented in a network-centric fashion, by dividing the APs into disjoint clusters as in Fig.~\ref{fig:figure1} (top-right). The APs in a cluster transmit jointly to the UEs residing in their joint coverage area, thus it is equivalent to deploying a conventional cellular network with distributed antennas in each cell. Despite the great theoretical gains, the 3GPP LTE (3rd Generation Partnership Project Long Term Evolution) standardization of CoMP-JT has not achieved much practical gains \cite{Fantini2016a}. This fact does not mean that the basic concept is flawed, but the network-centric approach may not be preferable.

Conversely, when the co-processing is implemented in a user-centric fashion, each user is served by coherent joint transmission from its selected subset of APs (\textit{user-specific cluster}), while all the APs that affect the user take its interference into consideration, as illustrated in Fig.~\ref{fig:figure1} (bottom-left). Hence, this approach eliminates the cell boundaries resulting in no inter-cell interference. Such transmission design, generalizable as \textit{user-specific dynamic cooperation clusters}~\cite{Bjornson2013d}, has been considered in MIMO cooperative networks~\cite{Tolli2008a,Garcia2010a,Kaviani2012a}, CoMP-JT~\cite{Baracca2012b}, cooperative small cells (\textit{cover-shifts})~\cite{Jungnickel2014b}, and C-RAN~\cite{Yuan2017c,Pan2018b}.

The combination of time-division duplex (TDD) Massive MIMO operation, dense distributed network topology, and user-centric transmission design creates a new concept, referred to as \textit{ubiquitous cell-free Massive MIMO}. To avoid preconceptions, we use the new \emph{``cell-free'' communication} terminology from \cite{Ngo2017b,Nayebi2015a} instead of prior terminology.
The word ``cell-free'' signifies that, at least from a user perspective, there are no cell boundaries during data DL transmission, but all (or a subset of) APs in the network cooperate to jointly serve the users in a user-centric fashion. The APs are connected via front-haul connections to central processing units (CPUs), which are responsible for the coordination. The CPUs are interconnected by back-haul (Fig.~\ref{fig:figure1}, bottom-right). In the UL, data detection can be performed locally at each AP, centrally at the CPU or partially first at each AP and then at the CPU. The UL spectral efficiency of cell-free Massive MIMO under four different levels of receiver cooperation are evaluated in~\cite{Bjornson2019a}. The full joint UL processing provides the best performance over any full or partial local processing, assuming the MMSE (minimum mean square error) combining is used. However, the more the CPU is involved in the processing, the higher the front-haul requirements are.

\begin{figure*}[b]\centering
	\includegraphics[width=.95\textwidth]{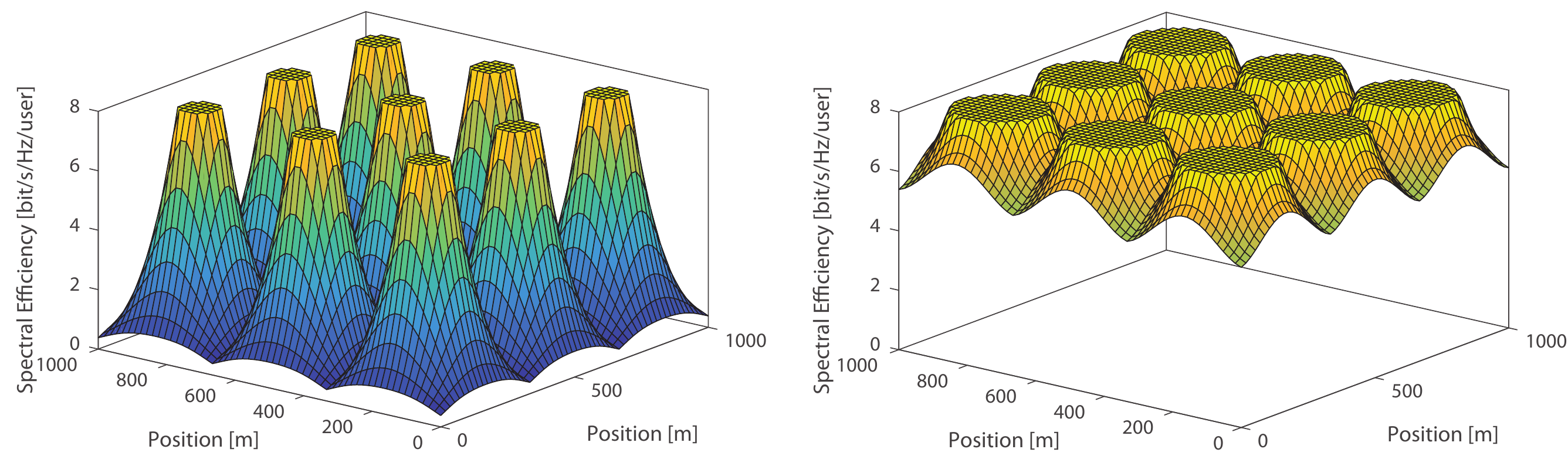} 
	  \caption{\csentence{Data coverage.}
      Left: cellular network. Right: cell-free Massive MIMO network. SE achieved by UEs at different locations in an area covered by nine APs that are deployed on a regular grid. Note that 8 bit/s/Hz was selected as the maximal SE, which corresponds to uncoded 256-QAM.}
	\label{fig:figure2}     
\end{figure*}

We stress that also in a ``cell-free'' network, we might have AP-specific synchronization and reference signals, which are important when accessing the network. 
More specifically, the UE initial access procedure in ``cell-free'' networks may follow the same principles as in LTE~\cite{Dahlman2013} or 5G-NR~\cite{Dahlman2018}, which are based on the cellular architecture. 
An \textit{inactive} UE first searches and then selects the best cell to camp on, by performing a so-called \textit{cell search and selection} procedure. By doing this, the UE acquires time and frequency synchronization with the selected cell and detects the corresponding \textit{Cell ID} as well as cell-specific reference signals, such as DMRS (demodulation reference signal) and CQI (channel quality indicator). 
Hence, the cellular architecture might be still underlying a ``cell-free'' network, and by the term ``cell-free'' we just mean that there are no cell boundaries created by the data transmission protocol in active mode.

\section{\normalsize System operation and resource allocation}

Ubiquitous cell-free Massive MIMO enhances the conventional (network-centric) CoMP-JT by leveraging the benefits of using Massive MIMO, i.e., high spectral efficiency, system scalability, and close-to-optimal linear processing. To give a first sense of the paradigm shift that cell-free Massive MIMO constitutes, Fig.~\ref{fig:figure2} shows the user performance at different locations in an area with nine APs: left figure shows that the SEs in a cellular network is poor at the cell edges due to strong inter-cell interference, while right figure shows that a cell-free network can avoid interference by co-processing over the APs and provide more uniform performance among the users. The SE is only limited by signal propagation losses.

\subsection{Ubiquitous Cell-Free Massive MIMO: The Scalable Way to Implement CoMP-JT}

The first challenge in implementing a cell-free Massive MIMO network is to obtain sufficiently accurate channel state information (CSI) so that the APs can simultaneously transmit (receive) signals to (from) all UEs and cancel interference in the spatial domain.
The conventional approach of sending DL pilots and letting the UEs feed back channel estimates is unscalable since the feedback load is proportional to the number of APs. Hence, frequency-division duplex (FDD) operation is not convenient, unless UL and DL channels are close enough in frequency to present similarities~\cite{Kim2018a}.
To circumvent this issue, we note that each AP only requires local CSI to perform its tasks~\cite{Bjornson2010c}. (Local CSI  refers to the channel between the AP and to each of the UEs.) 
This local CSI can be estimated from UL pilots, thus there is no need of exchanging CSI between the APs.
Local CSI is conveniently acquired in TDD operation since, when a UE sends a pilot, each AP can simultaneously estimate its channel to the UE. Hence, the overhead is independent of the number of APs. By exploiting channel reciprocity, the UL channel estimates can be also utilized as DL channel estimates, as in cellular Massive MIMO \cite{Marzetta2016a}.
Just like Massive MIMO is the scalable way to implement multi-user MIMO \cite{Marzetta2016a}, ubiquitous cell-free Massive MIMO is the scalable way to implement CoMP-JT.

In cell-free networks there are $L$ of geographically distributed APs that jointly serve a relatively smaller number $K$ of UEs: $L \gg K$. 
Cell-free Massive MIMO can provide ten-fold improvements in 95\%-likely SE for the UEs over a corresponding cellular network with small cells \cite{Ngo2017b,Nayebi2017a}. There are two key properties that explains this result.

\begin{figure*}\centering
	\includegraphics[width=.98\textwidth]{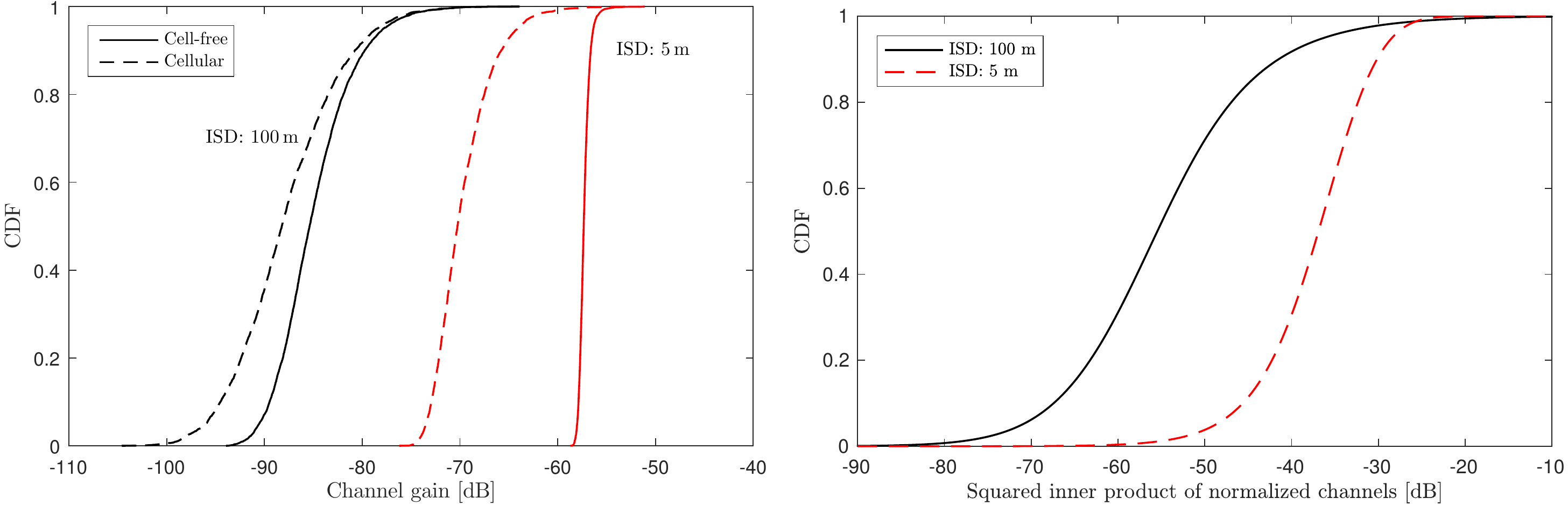} 
	  \caption{\csentence{Macro-diversity and favorable propagation.}
      Distribution of (left) the channel gain, and (right) the inner product of channel vectors in cell-free Massive MIMO. The simulation setup considers $2500$ single-antenna APs deployed on a square-grid with wrap-around and varying ISD. We consider independent Rayleigh small-scale fading and three-slope path-loss model from \cite{Ngo2017b}.}        
	\label{fig:figure3}      
\end{figure*}

\begin{figure*}[b]\centering
	\includegraphics[width=.7\textwidth]{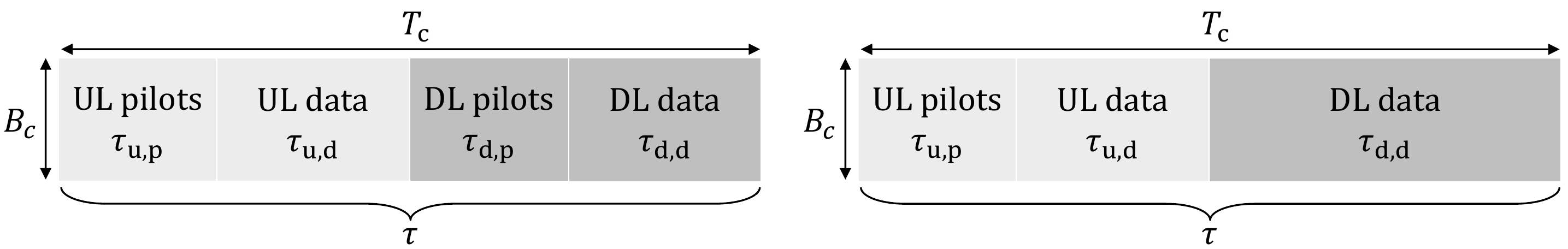}
		  \caption{\csentence{TDD frame structure.}
      The TDD frame with no pilot-based DL training (right) is used in cellular Massive MIMO, which can rely on channel hardening, while both options are on the table for cell-free Massive MIMO. Note that, guard intervals are not depicted since deducted from the coherence time interval.}        
	\label{fig:figure4}         
\end{figure*}

The first property is the increased macro-diversity. Fig.~\ref{fig:figure3} (left) illustrates this with single-antenna APs deployed on a square-grid with varying inter-site distance (ISD): 5, and 100\,m. The figure shows the cumulative distribution function (CDF) of the channel gain for a UE at a random position with channel vector $\mathbf{h} =[h_1 \, \ldots \, h_L]^T \in \mathbb{C}^{L}$, where $h_l$ is the channel from the $l$th AP. 
The channel gain is $\| \mathbf{h} \|^2$ in cell-free Massive MIMO and $\max_l |h_l|^2$ in a cellular network. With a large ISD, the UEs with the best channel conditions have almost identical channel gains in both cases, but the most unfortunate UEs gains 5\,dB from cell-free processing.
With a small ISD of 5\,m, which is reasonable for connected factory applications, all UEs obtain 5-20\,dB higher channel gain by the cell-free network.

The second property is \emph{favorable propagation}, which means that the channel vectors $\mathbf{h}_1,\mathbf{h}_2$ of any pair of UEs are nearly orthogonal, leading to little inter-user interference.
The level of orthogonality can be measured by the squared inner product 
\[
\frac{|\mathbf{h}_1^H\mathbf{h}_2|^2}{\| \mathbf{h}_1 \|^2 \| \mathbf{h}_2 \|^2}.\] 
A smaller value represents greater orthogonality. In a cellular network with single-antenna APs, $\mathbf{h}_1$ and $\mathbf{h}_2$ are scalars and thus the measure is one. Favorable propagation will, however, appear in cell-free Massive MIMO where $\mathbf{h}_1,\mathbf{h}_2 \in \mathbb{C}^{L}$, since the combination of small-scale and large-scale fading makes the large-dimensional channel vectors pairwise nearly orthogonal \cite{Chen2018b}. This is illustrated in Fig.~\ref{fig:figure3} (right), which shows the CDF of the orthogonality measure for two randomly located UEs. The inner product is very small for all the considered ISDs. Spatial correlated channels may hinder favorable propagation. In this case, proper user grouping and scheduling strategies can be implemented to reduce users' spatial correlation~\cite{Hajri2018a}. 

\subsection{TDD Protocol}

The TDD protocol recommended for cell-free Massive MIMO is illustrated in Fig.~\ref{fig:figure4}. Each AP estimates the UL channel from each UE by measurements on
UL pilots. By virtue of reciprocity, these estimates are also valid for the DL channels. Hence, the   pilot resource requirement is independent of the number of AP antennas and no UL feedback is needed.

After applying  precoding, each UE sees an  effective scalar channel. 
The UE needs to estimate the gain of this channel  to  decode its data. 
Note that  in cellular Massive MIMO, owing to \textit{channel hardening},
the UE may rely on knowledge of the average  channel gain for decoding \cite{Marzetta2016a}.
In cell-free Massive MIMO, in contrast, there is less hardening and 
DL effective gain estimation is desirable at the user 
\cite{interdonato2016dlpilot,Chen2018b}.
This estimate can either be obtained from DL pilots sent by the AP during a DL training phase \cite{interdonato2016dlpilot} (Fig.~\ref{fig:figure4}, left) or, potentially, blindly from the DL data transmission if there are no DL pilots (Fig.~\ref{fig:figure4}, right).
  
Fig.~\ref{fig:figure4} shows two possible TDD frame configurations, with and without DL pilot transmission. The configuration including the pilot-based DL training, depicted on the left in Fig.~\ref{fig:figure4}, consists of four phases: $(i)$ UL training; $(ii)$ UL data transmission; $(iii)$ Pilot-based DL training; and $(iv)$ DL data transmission. 
Fig.~\ref{fig:figure4}, on the right, illustrates the TDD frame without DL pilot transmission. This implies that, for data decoding, the UEs either rely on channel hardening or blindly estimate the DL channel from the data.   

The \textit{channel coherence interval} is defined as the time-frequency interval during which the channel can be approximately considered as static. It is determined by the propagation environment, UE mobility, and carrier frequency \cite{Marzetta2016a}.
The frequency-selectivity of the channel can be tackled by using OFDM (orthogonal frequency-division multiplexing), which transforms the  wideband channel into many parallel narrowband flat-fading channels~\cite{Marzetta2016a}. 
Alternatively, single-carrier modulation schemes can be used with similar performance~\cite{Pitarokoilis2012b, Filho2017}.
In regard to  handling channel frequency-selectivity, there is no conceptual difference between
cellular and cell-free Massive MIMO.

The TDD frame should be equal or shorter than the smallest coherence time among the active UEs. For simplicity, we herein assume it is equal.
Let $\tau = T_\mathrm{c}B_\mathrm{c}$ the length of TDD frame in samples, where $B_\mathrm{c}$ is the coherence bandwidth and $T_\mathrm{c}$ indicates the coherence time.  It is partitioned as $\tau = \tau_{\mathrm{u,p}} + \tau_{\mathrm{u,d}} + \tau_{\mathrm{d,p}} + \tau_{\mathrm{d,d}}$, where $\tau_{\mathrm{u,p}}, \tau_{\mathrm{u,d}}, \tau_{\mathrm{d,p}}$ and $\tau_{\mathrm{d,d}}$ denote the total number of samples per frame spent on transmission of UL pilots, UL data, DL pilots and DL data, respectively. Importantly, $\tau$ can be adjusted over time (by varying the values of $\tau_{\mathrm{u,p}}, \tau_{\mathrm{d,p}}, \tau_{\mathrm{d,d}}, \tau_{\mathrm{u,d}}$) to accommodate the coherence interval variation and the traffic load change. However, such frame reconfiguration should occur slowly to limit the amount of control signaling required by the resource re-allocation.   

The maximum number of mutually orthogonal pilots is upper-bounded by $\tau$. Hence, allocating a unique orthogonal pilot per user is physically impossible in networks with $K \geq \tau$, and either non-orthogonal pilots or pilot reuse are necessary. UEs that send non-orthogonal pilots (or share the same pilot) cause mutual interference that make the respective channel estimates correlated, a phenomenon known as \textit{pilot contamination}.

\subsection{Uplink Pilot Assignment}

To limit pilot contamination, efficient pilot assignment is important. We herein focus on uplink pilot assignment, but similar arguments are valid for downlink pilot assignment too~\cite{Interdonato2019b}.

Uplink pilot assignment is determined either locally at each AP, or centrally at the CPU. In the latter case, a message mapping the UE identifier to the pilot index is communicated to all the APs which forward it to the UEs. This UE-to-pilot mapping can be transmitted either in the broadcast control channel within the system information acquisition process or in the random access channel during the random access procedure.
Pilot assignment can be done in several ways: 

\begin{itemize}
\item Random pilot assignment: Each UE is randomly assigned one of the $\tau_{\mathrm{u,p}}$ mutually orthogonal pilots. This method requires no coordination, but there is a substantial probability that closely located UEs use the same pilot, leading to bad performance.

\item Brute-force optimal assignment: A search over all possible pilot sequences can be performed  to maximize a utility of choice, such as the max-min rate or sum rate. This method is optimal but its complexity grows exponentially with $K$.

\item Greedy pilot assignment~\cite{Ngo2017b} The $K$ UEs are first assigned pilot sequences at random.
Then this  assignment is iteratively improved by performing small
changes that increase the   utility.

\item Structured/Clustering pilot assignment~\cite{Attarifar2018a,Sabbagh2018a}: regular pilot reuse structures are adopted to guarantee that users sharing the same pilot are enough spatially separated, and ensure a marginal pilot contamination.  

\end{itemize}

\subsection{Power Control} 

Power control is important to handle the near-far effect, and protect UEs from strong interference.  The power control can be governed by the CPU, which tells the APs and UEs which power-control coefficients to use.
By using closed-form capacity bounds that only depend on the large-scale fading, the power control can be well optimized and infrequently updated, e.g., a few times per second. 

\begin{figure*}\centering
	\includegraphics[width=.6\textwidth]{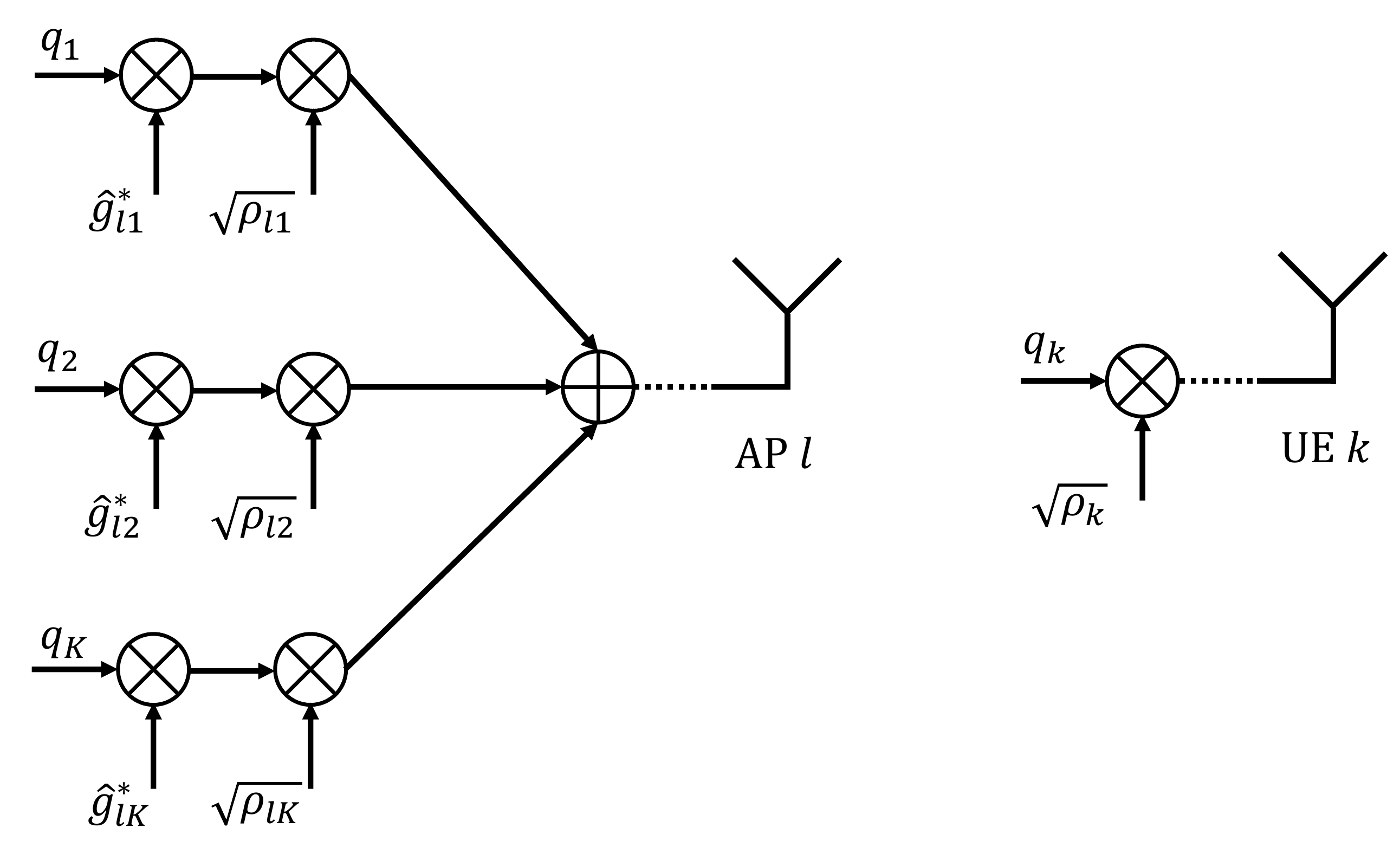} 
	  \caption{\csentence{Power control.}
      Processed signals at the $l$th AP (left) and the $k$th UE (right) with maximum ratio precoding/combining.}        
	\label{fig:figure4b}    
\end{figure*}

When maximum-ratio (MR) precoding is used, at AP $l$, the symbol intended for UE $k$, $q_k$, is first weighted by $\hat{g}_{lk}^\ast$ and $\sqrt{\rho_{lk}}$, where $\hat{g}_{lk}$ is the estimate of the channel from AP $l$ to UE $k$ and $\rho_{lk}$ is the power-control coefficient. The weighted symbols of all $K$ UEs will be then combined and transmitted to the UEs.
In the UL, at UE $k$, the corresponding symbol $q_k$ is weighted by a power-control coefficient $\sqrt{\rho_{k}}$ before transmission to the APs.
The block diagram that depicts the signal processing in the DL and the UL is shown in Fig.~\ref{fig:figure4b}.  

In general,  the power-control coefficients should be selected to maximize  a given performance objective. This objective may, for example, be the max-min rate or sum rate:

\begin{itemize}

\item Max-Min Fairness Power Control:
The goal of this power-control policy is to deliver the same rate to all UEs and maximizing that rate. In a large network, some UEs may have very bad channels to all APs, thus it is necessary to drop them from service before applying this  policy, otherwise the service will be bad for everyone.
As in cellular Massive MIMO, the max-min fairness power-control coefficients can be obtained efficiently by means of linear   and second-order cone optimization  \cite[Section IV-B]{Ngo2017b}.

\item Power Control with User Prioritization: 
The rate requirements are typically different among the UEs, which can be taken into account in the power-control policy. For instance, UEs that use real-time services or have more expensive subscriptions have higher priority.
The max-min fairness power control can be extended to consider weighted rates, where the individual weights represent the priorities. Minimum rate constraints can be also included.  

\item Power Control with AP Selection: Due to the path-loss, APs far away from a given UE will modestly contribute to its performance. AP selection is implemented by setting non-zero power-control coefficients to the APs designed to serve that UE.

\end{itemize}

Optimal power control is performed at the CPU. Centralized power-control strategies might jeopardize the system scalability and latency as the number of APs and UEs grows significantly. Simpler, scalable and distributed power-control policies, but providing decreased performance, are proposed in~\cite{Ngo2017b,Nayebi2017a,Interdonato2019a}.

To achieve good network performance, pilot assignment and power control can be performed jointly.

\section{\normalsize Practical Deployment Issues}

The cost and complexity of deployment, limited capacity of back/front-haul connections, and network synchronization are three major issues that need to be solved in a practical deployment.

\subsection{Radio Stripes System}

The cabling and internal communication between APs is challenging in practical cell-free Massive MIMO deployments. An appropriate, cost-efficient architecture is the \textit{radio stripe system}~\cite{patentRadioStripe}, presented next.

\begin{figure*}\centering
	\includegraphics[width=.9\textwidth]{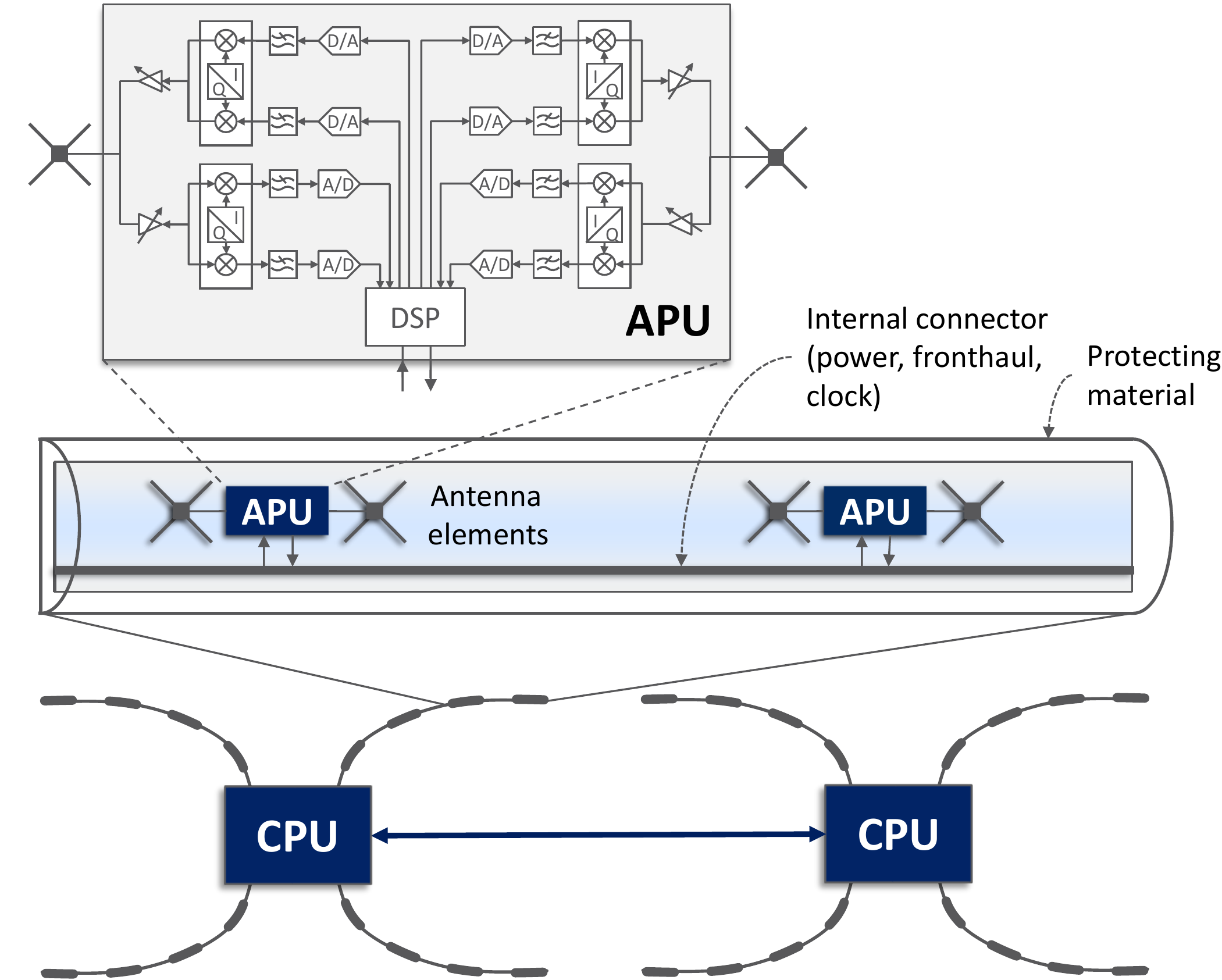}
	  \caption{\csentence{Radio stripe system design.}
      Each radio stripe sends/receives data to/from one or multiple CPUs through a shared bus (or internal connector), which also provides synchronization and power supply to each APU.}        
	\label{fig:figure5}        
\end{figure*}

In a radio stripe system, the antennas and the associated antenna processing units (APUs) are serially located inside the same cable, which also provides synchronization, data transfer, and power supply via a shared bus; see Fig.~\ref{fig:figure5}. More specifically, the actual APs consist of  antenna elements and circuit-mounted chips (including power amplifiers, phase shifters, filters, modulators, A/D and D/A converters) inside the protective casing of a cable or a stripe. Each radio stripe is then connected to one or multiple CPUs. A radio stripe embeds multiple antenna elements, where each antenna element effectively is an AP.  These APs could in turn cooperate phase-coherently. Hence, effectively a radio stripe constitutes a multiple-antenna AP. 
Moreover, depending on the carrier frequency, the multiple antennas can either be co-located (at higher frequencies the antenna elements are smaller) or distributed on the
radio stripe.
Since the total number of antennas is assumed to be large, the transmit power of each antenna can be very low, resulting in low heat-dissipation, small volume and weight, and low cost. Small low-gain antennas are used. For example, if the carrier frequency is 5.2\,GHz then the antenna size is 2.8\,cm, thus, the antennas and processing hardware can be easily fitted in a cable or a stripe.

\begin{figure*}[b]\centering
	\includegraphics[width=.7\textwidth]{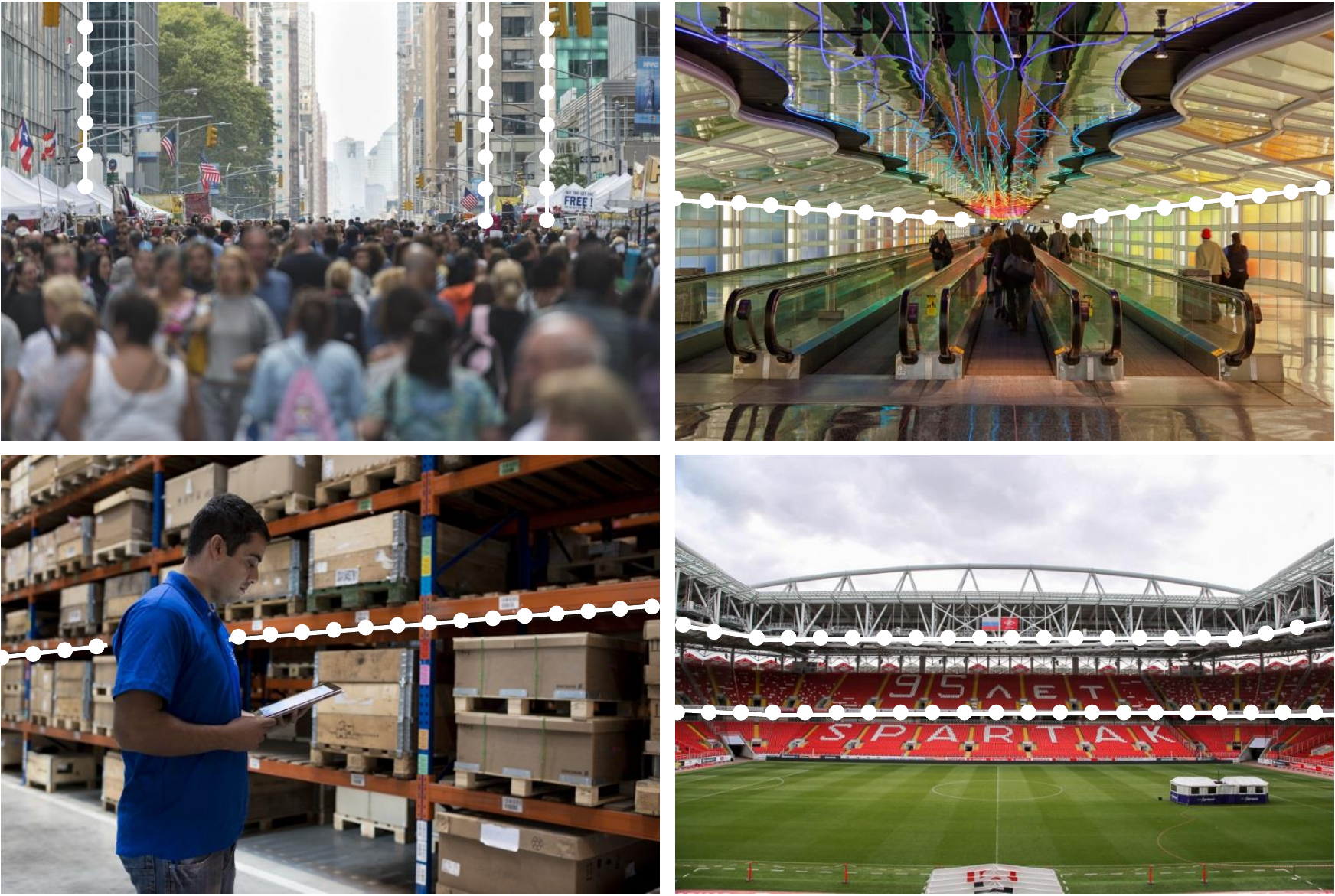}
  \caption{\csentence{Potential applications and deployment concepts.}
      Radio stripes, here illustrated in white, enable invisible installation in existing construction elements.}        
	\label{fig:figure6}  	      
\end{figure*}

The receive/transmit processing of an antenna is performed right next to itself. On the transmitter side, each APU receives up to $K$ streams of input data (e.g., one stream per UE, one UE with $K$ streams, or some other UE-stream allocation) from the previous APU via the shared bus. In each antenna, the input data streams are scaled with the pre-calculated precoding vector and the sum-signal is transmitted over the radio channel to the receiver(s). By exploiting   channel reciprocity, the precoding vector may be a function of the estimated UL channels. 
For example, if the conjugate of the estimated UL channel is used, MR precoding is obtained. This precoding requires no CSI sharing between the antennas.

On the receiver side, the received radio signal is multiplied with the combining vector previously calculated in the UL pilot phase. The output gives $K$ data streams. The processed streams are then combined with the data streams received from the shared bus and sent again on the shared bus to the next APU. More specifically, the $m$th APU sums its received data streams to the input streams from APU $m-1$ consisting of combined signals from APUs $1, \ldots, m-1$, for one or more UEs. This cumulative signal is then outputted to APU $m+1$. The combination of signals is a simple per-stream addition operation.

The radio stripe system facilitates a flexible and cheap cell-free Massive MIMO deployment. Cheapness comes from many aspects: $(i)$ deployment does not require highly qualified personnel. Theoretically, a radio stripe needs only one (plug and play) connection either to the front-haul network or directly to the CPU; $(ii)$ a conventional distributed Massive MIMO deployment requires a star topology, i.e., a separate cable between each  APs and a CPU, which may be economically infeasible. Conversely, radio stripe installation complexity is unaffected by the number of antenna elements, thanks to its compute-and-forward architecture. Hence, cabling becomes much cheaper. The star topology might be preferable from a performance perspective, but the cost of deployment of the front-haul network might be very high or even prohibitive. A way to efficiently use the long front-haul cables is to embed antenna elements into them, turning the cables into radio stripes. As a result, a star topology but with many radio stripes is obtained and the coverage improved; $(iii)$ maintenance costs are cut down as a radio stripe system offers increased robustness and resilience: highly distributed functionality offer limited overall impact on the network when few stripes being defected; $(iv)$ low heat-dissipation makes cooling systems simpler and cheaper. 

While cellular APs are bulky, radio stripes enable invisible installation in existing construction elements as exemplified in Fig.~\ref{fig:figure6}.
Moreover, a radio stripe deployment may integrate for example temperature sensors, microphones/speakers, or vibration sensors, and provide additional features such as fire alarms, burglar alarms, earthquake warning, indoor positioning, and climate monitoring and control.

\subsection{Front-haul and Back-haul Capacity}

While there is no need to share CSI between antennas, the CPUs must provide each APU with the data streams. The data is delivered from the core network via the back-haul and then forwarded to the APU over the front-haul; see Fig.~\ref{fig:figure5}. Similarly, the CPU receives the cumulative signals from its radio stripes over the front-haul and decodes them. The data will then be delivered to the core network over the back-haul.

The required front-haul capacity of a radio stripe is proportional to the number of simultaneous data streams that it supports at maximum network load. The required back-haul capacity of a CPU corresponds to the sum rate of the data streams that its radio stripes will transmit/receive at maximum network load. The way to limit these capacity requirements is to constrain the number of UEs that can be served per AP (e.g., radio stripe) and CPU. To avoid creating cell boundaries, a user-centric perspective must be used when selecting which subset of APs that serve a particular UE \cite{Ngo2017b,Buzzi2017b,Ngo2018a}, as illustrated on the bottom-left in Fig.~\ref{fig:figure1}.

Suppose a UE is alone in the network and all APs transmit to it with full power. Since the path-loss decays rapidly with the propagation distance, 95\% of the received power will originate from a subset of the APs, called the \emph{95\%-subset}.
When the ISD is large, as in a conventional cellular network, the 95\%-subset might only contain a handful of APs. As the ISD reduces (i.e., the number of APs per km$^2$ grows), the 95\%-subset is larger. This property can be used to limit the back-haul signaling. For example, it is shown in \cite{Ngo2017b} that only 10-20\% of the APs in the 1 km$^2$ area surrounding a UE belongs to the 95\%-subset.

\subsection{Synchronization}

To serve a UE by coherent joint transmission from multiple APs, the network infrastructure needs to be synchronized. The network might have an absolute time (phase) reference, but the APs are unsynchronized. This means that, effectively, the transmitter and receiver circuits of each AP have their own time references. The difference in time reference between the transmitter and receiver in a given AP represents the reciprocity calibration error. The difference in, say, transmitter time reference, between any pair of APs represents the synchronization error between these two APs. To limit the reciprocity and synchronization errors, a synchronization process needs to be applied at regular intervals. 

Suppose the transmitter of AP$_i$ has a clock bias of $t_i$ (i.e., its local time reference clock shows zero at absolute time $t_i$) and the receiver has a clock bias of $r_i$ (i.e., its clock shows zero at absolute time $r_i$). We propose a simple synchronization protocol that works as follows:
\begin{enumerate}
\item	At local time zero (absolute time $t_1$), AP$_1$ transmits a known pulse.
AP$_2$ receives this pulse at time $t_1-r_2$, according to its clock, and timestamps it with $\delta_{12}=t_1-r_2$. Similarly, AP$_3$ timestamps the pulse with $\delta_{13}=t_1-r_3$.

\item At its local time zero, AP$_2$ transmits a known pulse.
AP$_1$ timestamps the received pulse with its local reception time $\delta_{21}=t_2-r_1$. 
AP$_3$ timestamps it with $\delta_{23}=t_2-r_3$.

\item Finally, at its local time zero, AP$_3$ transmits a known pulse.
AP$_1$ timestamps this received pulse with $\delta_{31}=t_3-r_1$. 
AP$_2$ timestamps it with $\delta_{32}=t_3-r_2$.

\end{enumerate}
The quantities $\delta_{ij}$ are known from the measurements, but $t_1, r_1$, $t_2, r_2$, $t_3, r_3$ cannot be obtained from $\delta_{ij}$ since the corresponding linear equation system is singular.  However, the reciprocity and synchronization errors are easily recovered: 
\begin{align*}
t_1-r_1&=\delta_{12}+\delta_{31}-\delta_{32}, \\
t_2-r_2&=\delta_{21}+\delta_{32}-\delta_{31}, \\
t_3-r_3&=\delta_{31}+\delta_{23}-\delta_{21}, \\
t_1-t_2&=\delta_{13}-\delta_{23}, \\
t_1-t_3&=\delta_{12}-\delta_{32}, \\
t_2-t_3&=\delta_{21}-\delta_{31}.
\end{align*}
This enables synchronization between the three APs.

This synchronization method can be applied in a differential manner. Consider measurements $\delta_{ij}$ taken at a first point in time at which the biases are $t_1, r_1$, $t_2, r_2$, $t_3, r_3$, and then measurements  $\delta'_{ij}$ taken at a second point in time at which the biases are $t'_1, r'_1$, $t'_2, r'_2$, $t'_3, r'_3$. The application of the above method to $\delta'_{ij}-\delta_{ij}$ yields the evolution of clock biases, up to a drift that is common to the whole group.

Extension to synchronization between two groups is straightforward. Consider  two groups A and B, each group comprising three APs. The reciprocity and synchronization errors within each group may be calibrated through the above-described procedure. Each group will, however, have an unknown remaining clock bias.
Let $\delta^{\textrm{A,B}}_{ij} \triangleq t^{\textrm{A}}_i-r^{\textrm{B}}_j$ the time discrepancy measured at AP$_j$ in group B, following the known pulse transmission by AP$_i$ in group A. The inter-group synchronization error can be easily obtained by $t^{\textrm{A}}_i-t^{\textrm{B}}_j = \delta^{\textrm{A,B}}_{ik} - \delta^{\textrm{B,B}}_{jk}$.
Extensions to synchronization between more than two groups follows the same methodology as above. Note that, in a radio stripe system, groups of APs are sequential. Hence, synchronization is only required between a group and its neighbor.

\section{\normalsize Performance of Ubiquitous Cell-Free Massive MIMO}

We will analyze the anticipated performance, in terms of DL SE (bit/s/Hz/user), in two case studies of practical interest: $(i)$ an industrial indoor scenario, and $(ii)$ an outdoor piazza scenario. 
For both the cases we assume that the antenna elements, embedded in the radio stripes, implement MR precoding locally and no CSI is exchanged. Hence, each antenna element effectively acts as a single-antenna AP.  To evaluate the DL per-user SE, we use the closed-form expression for the DL capacity lower bound given in~\cite[Section III-B]{Ngo2017b}, which is valid for single-antenna APs implementing MR precoding and UEs relying on knowledge of the average channel gain for decoding. This closed-form expression is obtained under the assumption of independent Rayleigh fading channels, and accounts for channel estimation errors and interference from pilot contamination.

The two case studies differ in terms of propagation channel model, path-loss model, carrier frequency (which affects the antenna geometry), coverage requirements, and radio stripes layout deployment.    

\subsection{Industrial Indoor Scenario}

\begin{figure*}\centering
	\includegraphics[width=.97\textwidth]{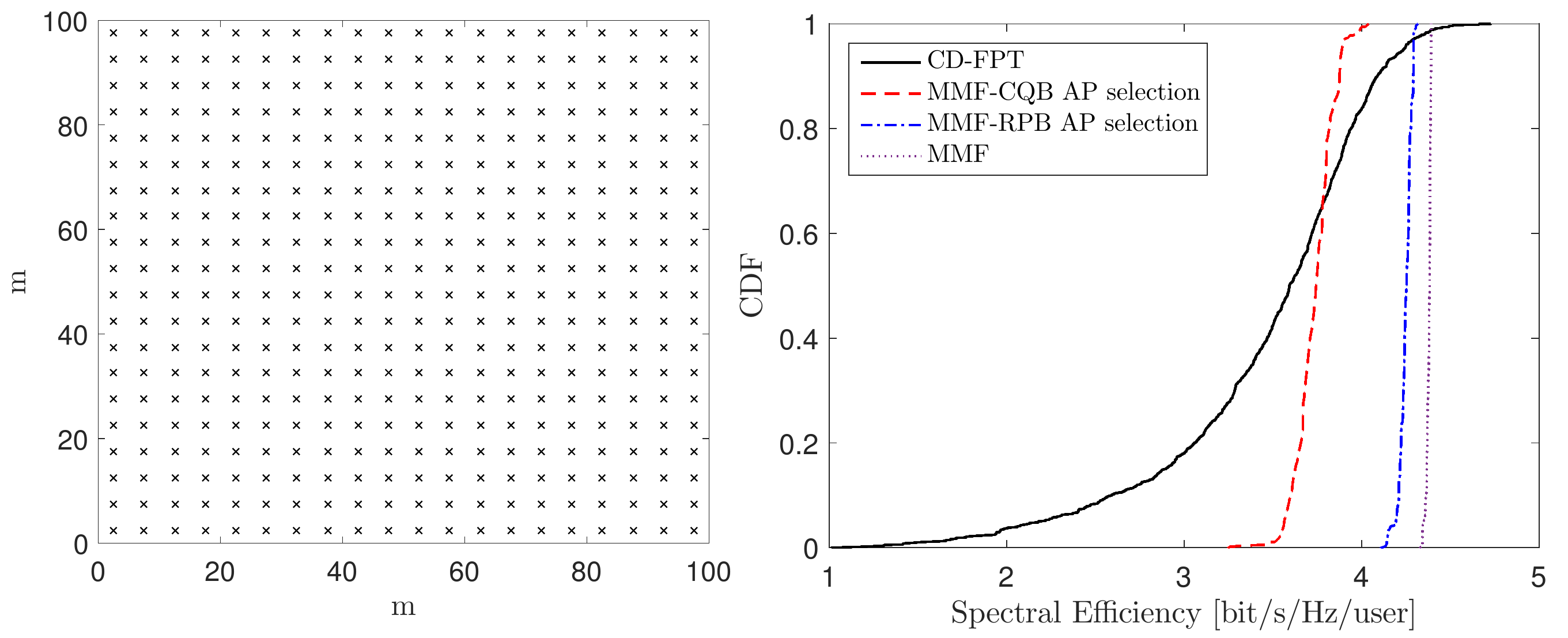}  
	  \caption{\csentence{Industrial indoor scenario.}
      Left figure illustrates the grid APs deployment. On the right, the CDF for the per-user SE, as defined in~\cite[Section III-B]{Ngo2017b}. In these simulations, we use the one-slope path-loss model defined in~\cite{tanghe2008industrialchannel}, with  reference distance $d_0=15$ m, path-loss at reference distance $\textrm{PL}(d_0)=70.28$, path-loss exponent $n=2.59$, and log-normal shadowing standard deviation $\sigma=6.09$. We choose $L=400$, $K=20$, bandwidth $B = 20$ MHz, and max per-AP radiated power 200 mW. The small-scale fading follows i.i.d.~Rayleigh distribution. We implement a wrap-around technique to simulate no cell boundaries.}        
	\label{fig:figure7}     
\end{figure*}

\begin{figure*}[b]\centering
	\includegraphics[width=.97\textwidth]{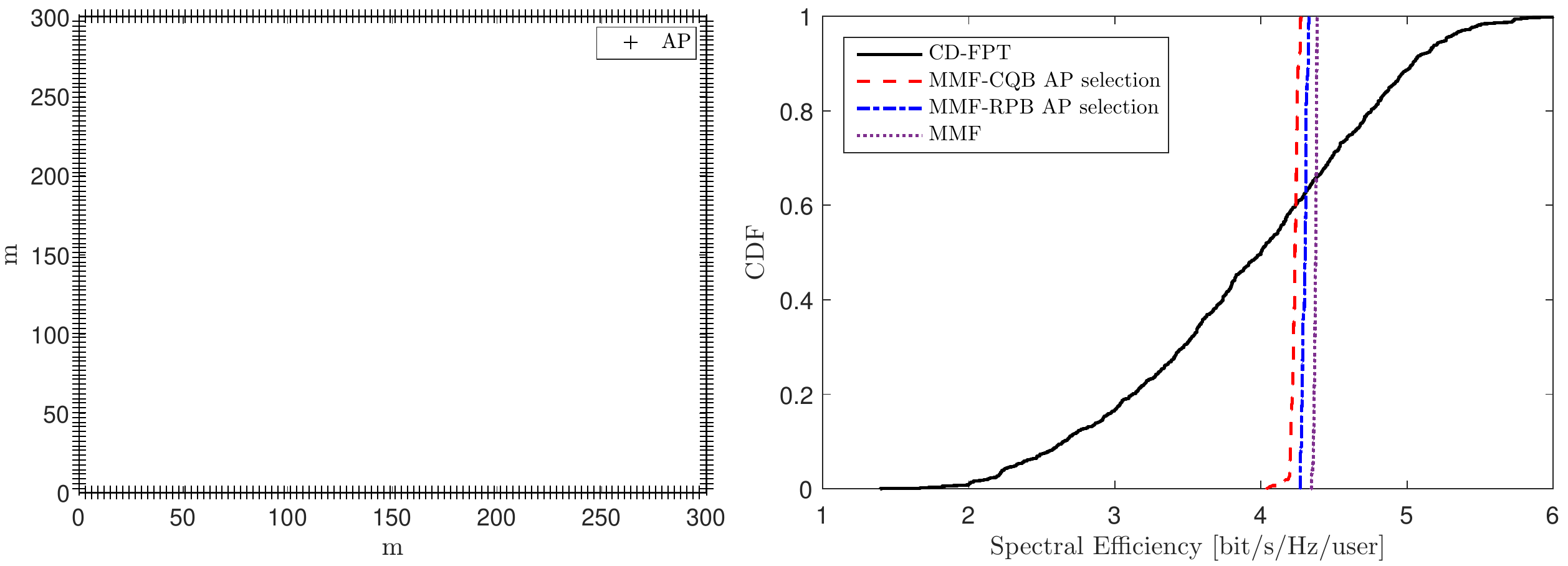} 
	  \caption{\csentence{Outdoor piazza scenario.}
     Left figure illustrates the APs deployed along the perimeter of the piazza. On the right, the CDF for the per-user SE, as defined in~\cite[Section III-B]{Ngo2017b}. In these simulations the large-scale fading is modeled as in~\cite{Ngo2017b}, assuming uncorrelated shadow fading. We choose $L=400$, $K=20$, bandwidth $B = 20$ MHz, and max per-AP radiated power 400 mW. The small-scale fading follows i.i.d.~Rayleigh distribution.}
	\label{fig:figure8}       
\end{figure*}

Ubiquitous coverage, low latency, ultra-reliable communication, and resilience are key for wireless communications in a factory environment. 
The flexible distributed cell-free architecture, with its macro-diversity gain and inherent ability to suppress interference, is   suitable to cope with the requirements of this scenario.

We consider the industrial indoor environment described in \cite{tanghe2008industrialchannel}: a 7-8\,m high building with metal ceiling and concrete floors and walls. The industrial inventory mainly consists of metal machinery. The radio stripes are deployed in an area of 100$\times$100 meters in such a way that 400 APs shape a 20$\times$20 regular grid, as shown in Fig.~\ref{fig:figure7} (left). The end-most antennas are 5\,m apart. They are placed at 6\,m above ground level, while the UE antenna height is 2\,m. The carrier frequency that we consider is 5200\,MHz, which is within the frequency band 5150-5825\,MHz adopted for application of indoor industrial wireless communications. Hence, a $\lambda/2$ antenna element (where $\lambda$ denotes the wavelength) has size 2.8\,cm.

The DL per-user SE and the impact of power control is shown in Fig.~\ref{fig:figure7} (right). We consider $K=20$ uniformly distributed UEs, mutually orthogonal UL pilots ($\tau_{\mathrm{u,p}}=K$), no DL training ($\tau_{\mathrm{d,p}}=0$), TDD frame length $\tau=200$ samples,  and four different DL power control settings, assuming a maximum per-AP radiated power of 200 mW:
\begin{enumerate}
\item CD-FPT: Channel-dependent full power transmission. All APs transmit with full power and the power-control coefficients for a given AP $l$ are the same for all $k=1, \ldots, K$. The power-control coefficient between AP $l$ and UE $k$ is $\rho_{lk} = \left(\sum_{k'=1}^K \gamma_{lk'}\right)^{-1}$, where $\gamma_{lk'}$ is the variance of the corresponding channel estimate $\hat{g}_{lk'}$;
\item MMF: Max-min fairness power control. All the APs are involved in coherently serving a given UE. The power control coefficients are chosen to maximize the minimum spectral efficiency of the network, as described in detail in~\cite[Section IV-B]{Ngo2017b}.
\item MMF-RPB AP selection~\cite{Ngo2018a}: Max-min fairness power control with received-power-based AP selection. Only a subset of APs serves a given UE $k$. The subset consists of the APs that contribute at least $\alpha\%$ (e.g., 95\%, as described before) of the power assigned to UE $k$. Mathematically, 
\[ \sum\limits_{l=1}^{|\mathcal{A}_k|} \frac{\varrho_{lk}}{\sum\nolimits_{j=1}^L \sqrt{\rho_{jk}}\gamma_{jk}} \geq \alpha\%, \] where $|\mathcal{A}_k|$ is the cardinality of the user-$k$-specific AP subset, and $\{\varrho_{1k}, \ldots, \varrho_{Lk}\}$ is the set of the coefficients $\varrho_{lk} \triangleq \sqrt{\rho_{lk}}\gamma_{lk}$  sorted in descending order.
\item MMF-CQB AP selection~\cite{Ngo2018a}: Max-min fairness power control with channel-quality-based AP selection. This method selects the APs with the best channel quality (largest large-scale fading coefficient) towards UE $k$ as follows
\[ \sum\limits_{l=1}^{|\mathcal{A}_k|} \frac{\bar{\beta}_{lk}}{\sum\nolimits_{j=1}^L \beta_{jk}} \geq \alpha\%, \]
where $\beta_{jk}$ is the large-scale fading coefficient of the channel between the $j$th AP and the $k$th UE, and $\{\bar{\beta}_{1k}, \ldots, \bar{\beta}_{Lk}\}$ is the set of the large-scale fading coefficients sorted in descending order.
\end{enumerate}
The AP selection in~\cite{Ngo2018a} is performed centrally at the CPU as full information on the channel large-scale fading coefficients to all users is needed. An alternative, distributed scheme is proposed in~\cite{Bursalioglu2018a}, where each AP autonomously decides   whether to participate in the service of a given user based on local pilot observations.

Max-min fairness power control doubles the 95\%-likely SE compared to the baseline CD-FPT case. Thanks to optimal power control, the radio stripe system can guarantee to each UE almost 4.5 bit/s/Hz. The performance with AP selection is also evaluated (dashed and dashed dotted lines). We can see that the SE reduction is minor if the RPB AP selection strategy is used,  while the CQB criterion leads to a 20\% reduction.
The performance gap is attributable to the cardinality of the corresponding AP subsets; on average, CQB uses 17\% of the APs and RPB uses 42\% of the APs.

\subsection{Outdoor Piazza Scenario}

Installations causing a big visual impact on the environment can be prohibited in areas like piazzas and historic places. In such a scenario, a radio stripe system can provide all the advantages previously described with an unobtrusive deployment. We consider a radio stripe system that covers a 300$\times$300\,meters square. The radio stripes are placed along the perimeter of the square at 9\,m height, for example,   on building   facades. 
There are 400 APs in total, as shown in Fig.~\ref{fig:figure8} (left).
We consider $K=20$ uniformly distributed UEs, mutually orthogonal UL pilots ($\tau_{\mathrm{u,p}}=K$), no DL training ($\tau_{\mathrm{d,p}}=0$), TDD frame length $\tau=200$ samples, and the same power-control policies as before. To deal with the large coverage area, we set the maximum per-APs radiated power to 400\,mW, and use the carrier frequency 2000\,MHz, which gives a $\lambda/2$ antenna element 7.5\,cm long. There is   actually no need for much higher transmit power in outdoor scenarios. The radiated power can be further lowered by adding more APs while guaranteeing the same coverage and performance.

The numerical results are shown in Fig.~\ref{fig:figure8} (right).
With max-min power control, we can provide a SE around 4.5\,bit/s/Hz/user, doubling the 95\%-likely SE compared to the baseline CD-FPT. 
Due to the AP deployment symmetry, the AP selection strategies perform almost equally well; CQB and RPB select around $1/3$ of the APs on average. The performance gap with respect the case with no AP selection is negligible, thus $2/3$ of the APs can be left out in the transmission towards a given UE.

\section{\normalsize Conclusion: Where there's a will, there's a way}

Cell-free Massive MIMO brings the best of two worlds: 
the macro-diversity from distributing many APs and the interference cancellation from cellular Massive MIMO. The TDD operation ensures system scalability and distributed processing as the channel estimation and precoding occur at each AP, thus no instantaneous CSI is exchanged over the front-haul. The user-centric data transmission suppresses the inter-cell interference and also contributes to reduce the front-haul overhead. Thanks to all these features, cell-free Massive MIMO succeeds where all the prior coordinated distributed wireless systems failed.   

While this article has outlined  the basic processing and implementation concepts,  many open issues remain, ranging from  communication theory to measurements and engineering efforts:

\begin{itemize}

\item \textbf{Power control:} While   (weighted) max-min power-control   is computationally tractable and provides uniform quality of service,
 it does not take actual traffic patterns into account. New power control algorithms are needed to   balance  fairness, latency, and network throughput, while permitting a distributed implementation.

\item \textbf{Distributed signal processing:} MR precoding/detection and synchronization can be distributed, as described earlier, but the data encoding/decoding must be carried out at one or multiple CPUs. The distribution of such signal processing tasks over the network is non-trivial, when looking for a good tradeoff between high rates and limited back-haul signaling.

\item \textbf{Resource allocation and broadcasting}: Scheduling, paging, pilot allocation, system information broadcast, and random access are basic functionalities that traditionally rely on 
a cellular architecture. New algorithms and protocols are needed for these tasks in cell-free networks.

\item \textbf{Channel modeling:} The performance analysis of cell-free networks have primarily considered Rayleigh fading channels. Practical channels are likely to contain a mix of line-of-sight and non-line-of-sight paths, and will likely differ substantially depending on the carrier frequency. Dedicated channel measurements followed by refined channel modeling are necessary to better understand the channel characteristics and fine-tune resource allocation algorithms.

\item \textbf{DL channel estimation:} Recent works~\cite{Chen2018b,Interdonato2019b} show that cell-free networks provide a low degree of channel hardening.  DL channel estimates, needed for data decoding, can either be obtained from DL pilots, which increases the pilot overhead, or by blind estimation techniques that uses the DL data. Dedicated algorithms for this estimation are needed.

\item \textbf{Compliance with existing standards:} The 5G standard is intended to be forward-compatible and only relies on cell-identities for the basic functionalities. It is likely that cell-free data transmission can be implemented in 5G, but further work in standardization and conceptual development is needed.

\item \textbf{Prototype development:} The step from a promising communication concept to a practical network requires substantial prototyping. The first working cell-free prototype may be pCell, where  \cite{Forenza2015a} describes a setup with 32 APs serving 16 UEs. Since every AP in a cell-free network has low cost and footprint, prototyping can be carried out using rather simple components. One can begin by demonstrating the synchronization and joint processing capabilities with a small number of APs in a limited area, and then continue with more APs and larger coverage area.

\end{itemize}


\begin{backmatter}

\section*{Abbreviations}

3GPP: 3rd generation partnership project;
AP: access point;
APU: antenna processing unit;
BS: base station;
CD-FPT: channel-dependent full power transmission;
CDF: cumulative distribution function;
CoMP-JT: coordinated multipoint with joint transmission;
CPU: central processing unit;
CQI: channel quality indicator;
CSI: channel state information;
DL: downlink;
DMRS: demodulation reference signal;
FDD: frequency-division duplex;
ISD: inter-site distance;
LTE: long term evolution;
MIMO: multiple-input multiple-output;
MMF: max-min fairness power control;
MMF-RPB: MMF with received-power-based AP selection;
MMF-CQB: MMF with channel-quality-based AP selection;
MMSE: minimum mean square error;
MR: maximum-ratio;
NR: new radio;
OFDM: orthogonal frequency-division multiplexing;
SE: spectral efficiency;
TDD: time-division duplex;
UE: user equipment;
UL: uplink.

\section*{Authors' contributions}
All authors have contributed to this research work, read and approved the final manuscript.

\section*{Authors' information}
This work was conducted when Giovanni Interdonato was with Ericsson Research (Ericsson AB), Link\"oping, Sweden. 

\section*{Funding}
This paper was supported by the European Union's Horizon 2020 research and innovation programme under grant agreement No 641985 (5Gwireless), and ELLIIT.

\section*{Availability of data and materials}
Data sharing is not applicable to this article as no datasets were generated or analysed during the current study.

\section*{Competing interests}
The authors declare that they have no competing interests.

\bibliographystyle{spphys} 
\bibliography{IEEEabrv,refs}      

\end{backmatter}
\end{document}